\pdfoutput=1
\documentclass[sigconf,openany]{acmart}
\renewcommand\footnotetextcopyrightpermission[1]{}
\usepackage{multirow}

\usepackage{soul}
\usepackage{enumerate}   
\usepackage{caption}
\usepackage{graphicx}
\let\emptyset\varnothing

\begin{document}

\title{PCAL: A Privacy-preserving Intelligent Credit Risk Modeling Framework Based on Adversarial Learning}

\author{Yuli Zheng, Zhenyu Wu, Ye Yuan, Tianlong Chen, and Zhangyang Wang}

\begin{abstract}
Credit risk modeling has permeated our everyday life. Most banks and financial companies use this technique to model their clients' trustworthiness. While machine learning is increasingly used in this field, the resulting large-scale collection of user private information has reinvigorated the privacy debate, considering dozens of data breach incidents every year caused by unauthorized hackers, and (potentially even more) information misuse/abuse by authorized parties. To address those critical concerns, this paper proposes a framework of \textit{\underline{p}rivacy-preserving \underline{c}redit risk modeling based on \underline{a}dversarial \underline{l}earning} (PCAL). PCAL aims to mask the private information inside the original dataset, while maintaining the important utility information for the target prediction task performance, by (iteratively) weighing between a privacy-risk loss and a utility-oriented loss. PCAL is compared against off-the-shelf options in terms of both utility and privacy protection. Results indicate that PCAL can learn an effective, privacy-free representation from user data, providing a solid foundation towards privacy-preserving machine learning for credit risk analysis.
\end{abstract}

\maketitle
\pagestyle{plain}

\section{Introduction}

\begin{figure*}[t]
	\vspace{-1em}
	\centering
	\includegraphics[width=0.8\linewidth]{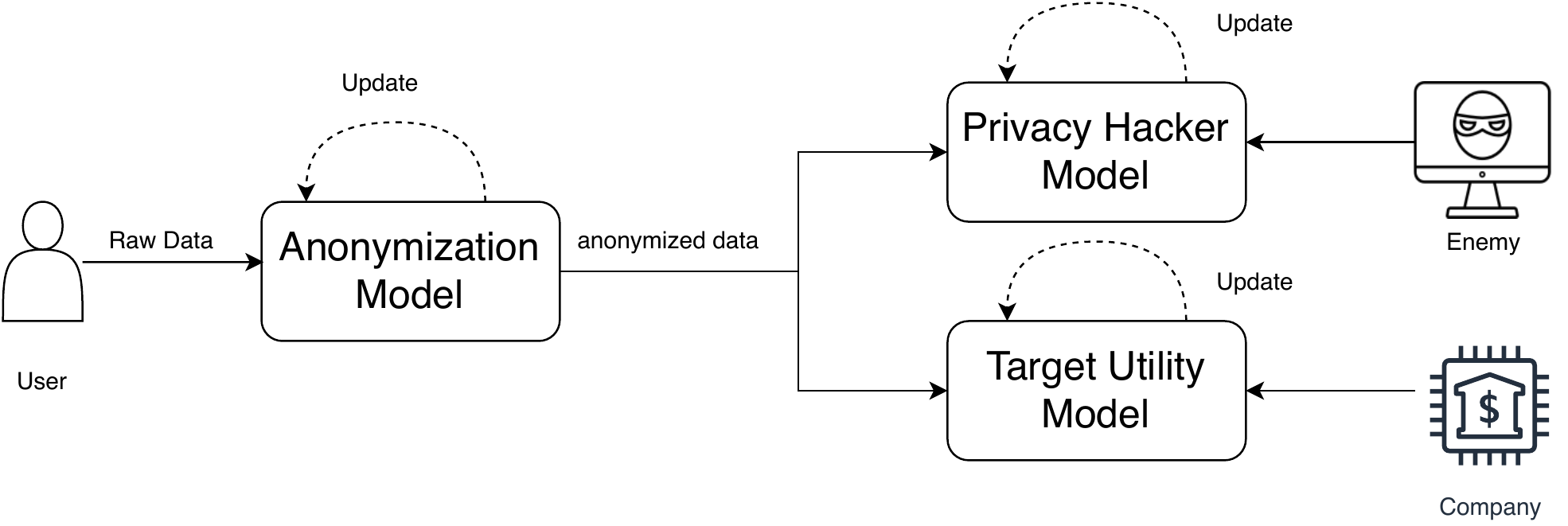}
	\caption{Overview of the proposed framework:  \textit{\underline{p}rivacy-preserving \underline{c}redit risk modeling based on \underline{a}dversarial \underline{l}earning} (PCAL).}
	\label{fig:frame}
	\vspace{-1em}
\end{figure*}

Credit scores -- numbers indicating the likelihood that a person will pay off his debts, such as FICO Score or Vantage Score, has now been the standard tool to evaluate millions of applicants' financial trustworthiness. Such a prevalence of credit scores has reinvigorated the privacy debate, since most of them require applicants to upload their personal information, including highly private and sensitive attribute such as annual income. Traditional cryptographic
solutions secure communication against unauthorized access from hackers.
However, they are not applicable to preventing authorized agents
(such as the bank analysts) from the unauthorized abuse of information, that
causes privacy breach concerns. The popular concept of differential privacy \cite{abadi2016deep} has
been introduced to prevent an adversary from gaining additional knowledge by
inclusion/exclusion of a subject, but not from gaining knowledge from released
data itself. In other words, an adversary can still accurately infer sensitive
attributes from any sanitized sample available, which does not violate any (proven) property of differential privacy.

This paper seeks to answer the question proposed by \cite{jennett2012privacy}: how to preserve your privacy while still allowing the banks to make the necessary predictions and analyses? 
At first glance, this question seems unsolvable: some of the privacy information itself is essential for the bank to make a decision or to give you a credit score, like annual income or account balance. We proposed a Privacy-preserving Credit risk modeling system based on Adversarial Learning(PCAL) to tackle this challenge. PCAL requires a learned \textit{masking} model deployed in the clients' end, which tries to anonymize the original data, and then pass its output -- anonymized data -- to the bank to let them analyze risk. The anonymization model directly optimizes the trade-off between analysis performance and associated privacy budget. The latter is quantified by monitoring the success rate of a (dynamic) privacy-hacking model.  
Strategies and evaluation protocols are also defined. Two experiments on real-life loan datasets demonstrate the effectiveness of PCAL.
\vspace{-0.5em}
\section{RELATED WORK}

\subsection{Credit Risk Modeling}
Plenty of researches have already applied machine learning techniques on credit risk modeling. \cite{khandani2010consumer} applied Classification And Regression Tree (CART) to construct forecasting models of consumer credit risk by combining customer transactions and credit bureau data. Deep belief networks were also leveraged to achieve results better than traditional credit scoring models such as logistic regression, multi-layer perceptron, and support vector machine \cite{luo2017deep}. 

Existing models aim to maximize predictive performance using all available information. However, not all data is essential for the desired utility, meanwhile, there is certainly privacy leak risk associated with (part of) the data. PCAL aims to strike a more balanced trade-off between the utility and the privacy, by learning to extract a task-related yet privacy-free representation from data.


\vspace{-0.5em}
\subsection{Privacy Preserving in Credit Modeling}
Privacy-preserving credit risk modeling is about protecting users' sensitive information and bridging the credibility gap between loan applicants and loan providers. 
\cite{fang2017survey} emphasizes the urgent need for privacy protection in risk modeling and calls for the laws and regulations to preserve users' data. \cite{jennett2012privacy} requires the loan providers to inform consent for sharing when collecting users' data and to collect user-permitted data only. However, many of the privacy-related data are essential to perform credit risk modeling. Also, this method requires perceived honesty and fairness from the loan provider, and can not prevent the potential privacy leak from happening. To the best of our knowledge, no previous project used any technically preserve privacy in credit risk modeling on data or model level, and we are the first that attempts to use a learning-based method to address this problem.

\vspace{-0.5em}
\subsection{Privacy Protection via Adversarial Learning}
Recently, a few learning-based methods have come into play to ensure better privacy protection. In \cite{raval2017protecting}, the authors proposed a game-theoretic framework between an obfuscator and an attacker, to hide visual secrets in the camera feed without significantly affecting the functionality of the target application. However, their method can only be applied to known single-attribute human-made privacy (a purposely attached QR code on a picture in their case) and cannot protect multiple privacy attributes and unknown inferred privacy information. \cite{wu2018towards,wang2019privacy, Wu_2019_ICCV} proposed a privacy-preserving learning framework for action recognition that prevents smart cameras at home from privacy invasion while allowing them to detect actions. PCAL is in many ways inspired by \cite{wu2018towards,wang2019privacy,Wu_2019_ICCV} while adapting many of their modules for the specific task of credit risk analysis.

\section{TECHNICAL APPROACH}

\subsection{Problem Definition}
Given training data $\mathcal{X}_t$ (users' financial data), our goal is to learn a representation that maintains a comparable target utility $\mathcal{T}$ (risk prediction performance) while preserving privacy by suppressing a privacy budget $\mathcal{P}$ (e.g., measuring how easily the sensitive attributes can be inferred from the learned features). Our goal can be mathematically expressed as below ($\lambda$ is a coefficient):
\begin{equation}
min_{f_T, f_A}L_T(f_T(f_A(X_t)), Y_T) + \lambda \Phi(f_A(X_t))
    \label{equation-one}
\end{equation}
Here, the $f_A$ is the anonymization model that learns a representation $f_A(\mathcal{X}_t)$ from the original data $\mathcal{X}_t$. The task model $f_T$ is to perform the target task $\mathcal{T}$ and predict label $Y_T$ (originally defined on $\mathcal{X}_t$). A standard loss function $L_T$ is used to evaluate the performance of the target task on the learned representation: the lower $L_T$, the better representation that can be used for risk analysis. A privacy leak metric $\Phi$ is used to evaluate the privacy budget $\mathcal{P}$: the larger $\Phi$, the higher privacy leak.

The objective hereby is to find a data anonymization model $f_A$ in a "Minimax Filter" ~\cite{hamm2017minimax} manner, such that 
\begin{enumerate}
\item the performance of $f_T$ is minimally affected, \\
\textit{i.e.}, $min_{f_T, f_A}L_T(f_T(f_A(X_t)), Y_T) \approx min_{f_T}L_T(f_T(X_t), X_t)$; 
\item the privacy leak metric $\Phi$ is minimized, \\
\textit{i.e.} $\Phi(f_A(X_t)) \ll \Phi(X_t)$.
\end{enumerate}

\subsection{Defining the Proxy for Privacy Cost}
The definition of privacy leak metric $\Phi$ is not simple. The $\Phi$ is a function to evaluate how much privacy leak exists in the learned representation $f_A(\mathcal{X}_t)$. 
Following \cite{wu2018towards}, we use an \textit{inference-based proxy} loss to empirically approximate $\Phi$, defined as below:
\begin{equation}
    \Phi(f_A(X_t) = - L_P(f_P(f_A(X_t)), Y_P)
\end{equation}
Here $Y_P$ is the privacy label in original data $\mathcal{X}_t$. $f_P$ is a privacy hacker model which tries to infer $Y_P$ from $f_A(X_t)$. $L_P$ is a loss function, which indicates the performance of privacy-hacking. Since the more easily a hack could happen, the more sensitive information the representation might contain, we reverse the sign to make sure that $\Phi$ increases as $L_P$ decreases.

This definition of $\Phi$ is based on the definition of one known $f_P$, but in reality, the hacker model is never pre-known, so it is not sufficient to suppress the success rate of only one hacker model. 

We therefore enhance our proxy, by considering a set of $M$ different privacy hacker models $\mathcal{H}_t=\{f_P^i \mid i=1,...,M\}$, used in training:
\begin{equation}
 \Phi(f_A(X_t) = - max_{f_P \in \mathcal{H}_t} L_P(f_P(f_A(X_t)), Y_P)
\label{equation-three}
\end{equation}
Our rationale lies in that the proper privacy protection should suppress every hacker model in $\mathcal{H}_t$, approximating the ideal requirement that it should withstand every possible hacker. For a solved $f_A$, the utility should be well sustained on $f_A(\mathcal{X}_t)$ using $f_T$, while no hacker (within or outside $\mathcal{H}_t$) can intrude into the protected privacy attributes. Implementation-wise, we \textbf{ensemble} the $M$ diverse models with winner-take-all, to form $\mathcal{H}_t$ in training.


\subsection{Overall Framework}
The PCAL (illustrated in Fig \ref{fig:frame}) uses the original data $\mathcal{X}_t$ as input. The main module to be learned, $f_A$, is used to anonymize the data while still preserving enough information for the predictive task utility. The anonymized representation is then passed as input to a target utility task model $f_T$ and a privacy hacker model $f_P$ at the same time. The entire model is trained using the function defined in (\ref{equation-one}), in an adversarial learning fashion. $f_A$ expects to find an optimal task-aware yet privacy-removing transformation. 


During adversarial training, we find a \textbf{restarting} technique useful to avoid trapping in bad local minima: when $\Phi(f_A(\mathcal{X}_t)$ is observed to not decrease anymore, we reset the weights in $f_P$ to random, while keeping other modules with their current weights unchanged, and continue training from there. The performance can often keep improving afterward.




Our last question to address is \textit{how to evaluate}. We need to ensure our anonymized data is robustly reliable against any unseen privacy hackers. We adopt the empirical evaluation protocol in \cite{wu2018towards}: given evaluation data $\mathcal{X}_e$, we first \underline{re-sample} a different set of $N$ models, $\mathcal{H}_{e}=\{f_P^i \mid i=1,...,N\}$ that are disjoint with $\mathcal{H}_t$; we then train each of them to predict the privacy attributes from the anonymized representations and test their generalization. The best prediction performance empirically indicates the ``worst case defense performance" of the learned representation, and will be used to evaluate the privacy protection effectiveness. Specifically,
\begin{align}
     \Phi(f_A(X_e) = max_{f_P \in \mathcal{H}_e} L_P(f_P(f_A(X_e)), Y_P), \text{where}\ \mathcal{H}_e \cap \mathcal{H}_t=\emptyset.
\end{align}
\section{EXPERIMENTS}
\subsection{Datasets and Implementation Details}

\begin{table}[]
\label{tablea}
\begin{tabular}{|c|c|l|l|l|l|}
\hline
\multicolumn{2}{|c|}{}              & \multicolumn{1}{c|}{WP}    & \multicolumn{1}{c|}{SP}    & \multicolumn{1}{c|}{PCAL}  & \multicolumn{1}{c|}{UP} \\ \hline
\multicolumn{2}{|c|}{Loan Decision Accuracy} & \multicolumn{1}{c|}{77.14} & \multicolumn{1}{c|}{57.40} & \multicolumn{1}{c|}{95.29} & 97.39                   \\ \hline
\multirow{8}{*}{\begin{tabular}[c]{@{}c@{}}Privacy \\ Attacking \\ Result \\ ($R^{2}$) \end{tabular}} &
  SVR &
  \multicolumn{1}{c|}{0.68} &
  \multicolumn{1}{c|}{0.28} &
  \multicolumn{1}{c|}{0.29} &
  \multicolumn{1}{c|}{1} \\ \cline{2-6} 
            & RFR                   & \multicolumn{1}{c|}{0.62}  & \multicolumn{1}{c|}{0.38}      & \multicolumn{1}{c|}{0.29}      &    \multicolumn{1}{c|}{1}                       \\ \cline{2-6} 
            & ElasticNet            & \multicolumn{1}{c|}{0.58}     & \multicolumn{1}{c|}{0.27}     & \multicolumn{1}{c|}{0.27}     &     \multicolumn{1}{c|}{1}                    \\ 
            \cline{2-6} 
            & Wide Net              &   \multicolumn{1}{c|}{0.40}                         &    \multicolumn{1}{c|}{0.18}   &    \multicolumn{1}{c|}{0.28}                        &   \multicolumn{1}{c|}{1}                 \\ \cline{2-6} 
            & Narrow Net            &      \multicolumn{1}{c|}{0.37}                      &     \multicolumn{1}{c|}{0.23}                       &    \multicolumn{1}{c|}{0.07}                        &        \multicolumn{1}{c|}{1}                 \\ \cline{2-6} 
            & Shallow Net           &     \multicolumn{1}{c|}{0.46}                       &    \multicolumn{1}{c|}{0.19}                        &     \multicolumn{1}{c|}{0.18}                       &        \multicolumn{1}{c|}{1}                 \\ \cline{2-6} 
            & Deep Net              &   \multicolumn{1}{c|}{0.49}                         &   \multicolumn{1}{c|}{0.20}                         &      \multicolumn{1}{c|}{0.39}                      &       \multicolumn{1}{c|}{1}                  \\ \cline{2-6} 
            & Standard Net          &    \multicolumn{1}{c|}{0.38}                        &     \multicolumn{1}{c|}{0.31}                       &    \multicolumn{1}{c|}{0.07}                        &      \multicolumn{1}{c|}{1}                   \\ \hline
\end{tabular}
	\caption{Loan Decision and Privacy Attacking Performance on BLS}
	\label{tablea}
\end{table}

\begin{table}[]
\begin{tabular}{|c|c|l|l|l|l|}
\hline
\multicolumn{2}{|c|}{}              & \multicolumn{1}{c|}{WP}    & \multicolumn{1}{c|}{SP}    & \multicolumn{1}{c|}{PCAL}  & \multicolumn{1}{c|}{UP} \\ \hline
\multicolumn{2}{|c|}{Loan Decision Accuracy} & \multicolumn{1}{c|}{77.14} & \multicolumn{1}{c|}{57.40} & \multicolumn{1}{c|}{95.29} & 97.39                   \\ \hline
\multirow{8}{*}{\begin{tabular}[c]{@{}c@{}}Privacy \\ Attacking \\ Result \\ ($R^{2}$) \end{tabular}} &
  SVR &
  \multicolumn{1}{c|}{0.48} &
  \multicolumn{1}{c|}{0.17} &
  \multicolumn{1}{c|}{0.19} &
  \multicolumn{1}{c|}{1} \\ \cline{2-6} 
            & RFR                   & \multicolumn{1}{c|}{0.38}  & \multicolumn{1}{c|}{0.16}      & \multicolumn{1}{c|}{0.17}      &      \multicolumn{1}{c|}{1}                   \\ \cline{2-6} 
            & ElasticNet            & \multicolumn{1}{c|}{0.29}     & \multicolumn{1}{c|}{0.13}     & \multicolumn{1}{c|}{0.18}     &       \multicolumn{1}{c|}{1}
            \\ \cline{2-6} 
            & Wide Net              &   \multicolumn{1}{c|}{0.17}                         &     \multicolumn{1}{c|}{0.08}                       &     \multicolumn{1}{c|}{0.06}                       &                   \multicolumn{1}{c|}{1}      \\ \cline{2-6} 
            & Narrow Net            &     \multicolumn{1}{c|}{0.06}                       &                 \multicolumn{1}{c|}{0.06}           &                    \multicolumn{1}{c|}{0.05}        &                \multicolumn{1}{c|}{1}         \\ \cline{2-6} 
            & Shallow Net           &          \multicolumn{1}{c|}{0.29}                  &            \multicolumn{1}{c|}{0.09}                &              \multicolumn{1}{c|}{0.09}              &               \multicolumn{1}{c|}{1}          \\ \cline{2-6} 
            & Deep Net              &           \multicolumn{1}{c|}{0.19}                 &           \multicolumn{1}{c|}{0.13}                &          \multicolumn{1}{c|}{0.11}                  &            \multicolumn{1}{c|}{1}             \\ \cline{2-6} 
            & Standard Net          &          \multicolumn{1}{c|}{0.17}                  &            \multicolumn{1}{c|}{0.12}                &              \multicolumn{1}{c|}{0.10}              &               \multicolumn{1}{c|}{1}          \\ \hline
            
\end{tabular}
	\caption{Loan Decision and Privacy Attacking Performance on LCL}
	\label{tableb}
\end{table}

We evaluate PCAL on two different datasets. The BLS dataset\cite{bankloan} with 50k records from loan borrowers, consists of privacy-sensitive information (\textit{i.e.} name, income, and account balance) and the risk evaluation (decision to the loan application) by professional bankers. The Lending Club Loan dataset\cite{lendingclubloan} contains an 890k record of loan issued between 2007-2015. 

Here we define the target utility task $\mathcal{T}$ to be loan decision and the privacy budget task $\mathcal{P}$ to be the inferring of one or multiple sensitive attributes (annual income, total balance). The goal here is to learn the masked representation that risk-related information is kept while privacy-relevant attributes are removed. The results is shown in table \ref{tablea} \& \ref{tableb}.



We utilized a three-layer fully-connected neural network for each of the choices of anonymization model $f_A$, task model $f_T$, and privacy model $f_P$. In the ensemble setting, we use five fully-connected neural networks with different width and depth to ensemble $f_P$, trained with MSE losses $L_T$ and $L_P$.

\subsection{Evaluation and Baselines} 
The privacy-preserving performance is evaluated by regressing the privacy-sensitive attribute values. We use r-squared error to show privacy-hacking result.
To cover as much as possible privacy-hacking models in $\mathcal{R}$, we manually selected eight unseen models, including Support Vector Regression\cite{drucker1997support} (SVR), Random Forest Regression\cite{liaw2002classification} (RFR), ElasticNet Regression\cite{Zou05regularizationand} and five neural networks with different widths and depths (Wide Net, Narrow Net, Shallow Net, Deep Net, and Standard Net). We tested the masked representations with all eight hackers and these models come to a unanimous conclusion. The results is shown in table \ref{tablea} \& \ref{tableb}.

Due to the absence of peer methods, we design the following baselines to compare with PCAL, to demonstrate the superior utility-privacy trade-off it can achieve:
\begin{enumerate}[(i)]
\item \textbf{Weak Protection (WP)}: removing only the explicit privacy attributes from the original data.
\item \textbf{Strong protection (SP)}:
removing all the attributes, whose correlations with the privacy part are above a threshold (0.4), and only use those weakly related data (\textit{e.g.} loan purpose, number of open accounts).
\item \textbf{Unprotected (UP)}: 
Directly using the unprotected raw data.
\end{enumerate}

\subsection{Results and Analysis} 
Table \ref{tablea} \& \ref{tableb} report the performance obtained by 8 models on those methods. We summarize observations as below:
\begin{enumerate}
\item PCAL has clear advantages over the naive WP and SP in the task accuracy. This makes sense since simply removing privacy attributes (either in a weak or a strong sense) causes inevitable damage to the target task performance.

\item PCAL maintains a competitive task utility performance \textit{w.r.t.} UP, while being much superior in protecting privacy (almost as well as SP), showing the most ideal trade-off.

\end{enumerate}
\section{Conclusions}
To the best of our knowledge, this work is the first attempt to design an adversarial learning framework for preserving privacy in credit risk analysis. We hope PCAL can help in part address the privacy leak concern for financial company clients.


\end{document}